\documentclass[hyper]{JHEP3}

\input{epsf}
\usepackage{epsfig}
\usepackage{amssymb}
\usepackage{amsfonts}
\usepackage{amsbsy}

\def\IR{{\mathbb{R}}}

\def\CN {{\cal N}}

\def\one{{\hbox{ 1\kern-.8mm l}}}

\title{How many black holes fit on the head of a pin?\,\footnote{Fourth prize in the Gravity Research Foundation
Essay competition 2007.}}

\author{Frederik~Denef$^1$ and Gregory~W.~Moore$^2$\\
$^1$ Instituut voor Theoretische Fysica, KU Leuven, \\
Celestijnenlaan 200D, B-3001 Leuven, Belgium \\
\\
$^2$ NHETC and Department of Physics and Astronomy,
Rutgers University,\\
Piscataway, NJ 08855--0849, USA\\
\\
{\tt frederik.denef@fys.kuleuven.be, gmoore@physics.rutgers.edu} }

\abstract{The Bekenstein-Hawking entropy of certain black holes can
be computed microscopically in string theory by mapping the elusive
problem of counting microstates of a strongly gravitating black hole
to the tractable problem of counting microstates of a weakly coupled
D-brane system, which has no event horizon, and indeed comfortably
fits on the head of a pin. We show here that, contrary to widely
held beliefs, the entropy of spherically symmetric black holes can
easily be dwarfed by that of stationary multi-black-hole
``molecules'' of the same total charge and energy. Thus, the
corresponding pin-sized D-brane systems do not even approximately
count the microstates of a single black hole, but rather those of a
zoo of entropically dominant multicentered configurations.}

\begin{document}

Explaining the microscopic origin of the entropy of black holes has
been a longstanding problem \cite{Wald,Damour:2004kw}. Impressive
progress was made, within the context of string theory, by
Strominger and Vafa, who accounted for the entropy of certain
supersymmetric five dimensional charged black holes in terms of
D-brane microstates \cite{Strominger:1996sh}. This was extended to
certain four dimensional supersymmetric black holes in
\cite{Maldacena:1997de,Vafa:1997gr}.

In this essay, we revisit some of the assumptions made in these
derivations, in particular the crucial and rather stunning assertion
that by tuning the string coupling to zero one can fit a black hole,
without loss of entropy, on the head of a pin.  We do not dispute
this fact --- on the contrary, we will show that quite a bit more
fits on that pin, namely supersymmetric multi-centered black hole
bound states, and we will demonstrate that these can actually dwarf
the entropy of single centered black holes. Although we do not take
the titular question as seriously as the medieval scholastics
allegedly took theirs \cite{Duns}, the answer has some tangible
implications for the program of reproducing statistically the
entropy of black holes.

We will focus on black holes in four dimensional $\CN=2$
supersymmetric theories. The basic idea underlying the microscopic
entropy computation is simple. The weighted number of supersymmetric
one-particle states with a given conserved charge $\Gamma$ is given
by the Witten index
\begin{equation} \label{indexdef}
 \Omega(\Gamma;\tau_\infty) = {\rm Tr}_{\, \Gamma} \, (-1)^{2 J_3} \, (J_3)^2 \, e^{-\beta H},
\end{equation}
where $J_3$ is the 3-component of the angular momentum, $H$ is the
energy above the supersymmetric lower bound, the trace is over the
subsector of the Hilbert space of states with charge $\Gamma$, and
$\tau_\infty$ denotes the values at spatial infinity of certain
scalars parametrizing the vacua of the theory. The index has the
remarkable property of being independent of $\beta$ and other
couplings of the theory since nonsupersymmetric (i.e.\ $H>0$) states
cancel in bose-fermi pairs, allowing it to be computed exactly in a
suitably controlled regime \cite{Witten:1982df}.

Now, in four dimensional string  compactifications, in the limit of
weak  coupling  and small internal metric curvatures, the Hilbert
space of  charged supersymmetric one particle ground states is well
understood: it is given by quantizing the moduli space of
supersymmetric configurations of D-branes wrapping various cycles of
the Calabi-Yau compactification manifold $X$,  and sitting at a
\emph{single} point in the noncompact spacetime --- that point being
the titular pinhead. On the other hand, the four dimensional low
energy effective supergravity theory has supersymmetric black hole
solutions carrying identical charges. Identifying these as the
classical strong gravitational coupling description of the weakly
coupled D-brane states and exploiting the deformation invariance of
the index, one thus obtains an indirect way to compute the
microcanonical statistical entropy of a black hole by counting the
corresponding supersymmetric D-brane ground states
\cite{Strominger:1996sh}.

Although this approach has been spectacularly successful in many
cases, leading to exact quantitative agreement with the macroscopic
Bekenstein-Hawking formula for the entropy, one could raise several
objections. We focus on one of them, which we believe is by far the
most significant one, in light of results we recently reported in
\cite{Denef:2007vg}.

The problem originates in the fact that the trace (\ref{indexdef})
runs over \emph{all} one particle states of charge $\Gamma$, not
just those corresponding to spherically symmetric black holes. For
example, there could be additional particles orbiting   the black
hole, or entire galaxies for that matter --- any state with total
charge $\Gamma$ should \emph{a priori} be included. Hence one could
wonder if the identification of the index with the ground state
degeneracy of a single, spherically symmetric black hole is really
justified.

This objection is usually not considered to be  serious, since
orbiting galaxies, being non-supersymmetric, just drop out of the
trace (\ref{indexdef}). Only supersymmetric configurations need be
considered. When \cite{Strominger:1996sh,Maldacena:1997de} appeared,
the only known   supersymmetric solutions were spherically symmetric
black holes \cite{Ferrara:1995ih,Strominger:1996kf}. Moreover, from
known results about black hole physics, it seemed intuitively
obvious that the most entropic minimal energy supergravity solution
should be a single black hole. It thus made perfect sense to
identify the leading D-brane entropy with the leading single
centered black hole entropy.

Subsequently, however, more general supersymmetric solutions were
discovered
--- stationary, multicentered, ``molecular'' black hole bound states
\cite{Behrndt:1997ny,Denef:2000nb,LopesCardoso:2000qm,Bates:2003vx}.
These are completely determined by harmonic functions
\begin{equation} \label{harmfunct}
 H^\Lambda(\vec x):= \sum_i \frac{p_i^\Lambda}{|\vec x - \vec{x}_i|} +
 h^\Lambda, \qquad H_\Lambda(\vec x) := \sum_i \frac{q_{i,\Lambda}}{|\vec x - \vec{x}_i|} +
 h_\Lambda,
\end{equation}
where $\vec{x}_i$ is the position of the $i$-th center, $\Gamma_i :=
(p^\Lambda_i,q_{\Lambda,i})$ its (magnetic, electric) charges with
respect to the $U(1)^{n+1}$ gauge group labeled by
$\Lambda=0,\ldots,n$, $(h^\Lambda,h_{\Lambda})$ are constants
determined by the total charge $\Gamma = \sum_i \Gamma_i$ and
$\tau_\infty$, and we work in units with Newton's constant $G_N
\equiv 1$.

Remarkably, given $H(\vec x)$, the   explicit solution for
\emph{all} fields can   be obtained from   one ``master'' function
$S$ defined on the $2n+2$ dimensional charge space and
 scaling as $S(\lambda \Gamma) = \lambda^2
S(\Gamma)$ \cite{Bates:2003vx}. In particular, the metric is
\begin{equation} \label{metricdef}
 ds^2 = -\frac{\pi}{S(H)} (dt + \omega)^2 + \frac{S(H)}{\pi} d \vec{x}^2
\end{equation}
where $\omega = \omega_i(\vec x) \, dx^i$ solves
\begin{equation} \label{domegaeq}
 d \omega = \langle *dH,H \rangle,
\end{equation}
with $\langle A,B \rangle := A^\Lambda B_\Lambda - A_\Lambda
B^\Lambda$ the duality invariant product and $*$ the Hodge star on
$\IR^3$.

The scaling homogeneity of $S$ together with (\ref{metricdef})
  implies that the area of the horizon of the $i$-th
center at $\vec x = \vec x_i$ is given by $A_i = 4 S(\Gamma_i)$, and
hence the corresponding Bekenstein-Hawking entropy equals
$S(\Gamma_i)$. Therefore $S$ is the single centered supersymmetric
entropy function, which in turn is completely determined by the
topological data of the string compactification.
%
%As an example, for a toroidal compactification, $S$ is given by the
%square root of the unique duality invariant quartic polynomial of
%the charges.
%
In order for the solution to exist, $H(\vec x)$ must remain within
the domain where the function $S$ is positive.

Equation (\ref{domegaeq}) can be solved provided the integrability
condition $\langle \Delta H, H \rangle = 0$ is satisfied. Since
$\Delta H \sim \sum_i \delta^3(\vec x - \vec x_i) \, \Gamma_i$, the
$\vec x_i$ are constrained and hence  these configurations are
typically genuine bound states, in the sense that one cannot move
the centers away from each other without input of energy. For
example for two centered solutions one gets the constraint
\begin{equation} \label{towcentersep}
 |\vec x_1 - \vec x_2| = - \frac{\langle \Gamma_1,\Gamma_2 \rangle}{\langle \Gamma_1,h
 \rangle}.
\end{equation}
Since $h$ depends on  $\tau_\infty$ and the right hand side must be
positive, the existence of these black hole bound states depends on
the choice of vacuum.

Thus, we face a problem: the index (\ref{indexdef})  receives
nonvanishing contributions not just from a single black hole, but,
in general, \emph{also} from a  disturbingly complex zoo of
multicentered black hole bound states with the same total charge
$\Gamma$.  All of these collapse to a single D-brane in the weak
string coupling limit \cite{Denef:2002ru}. This answers the titular
question definitively as ``Many, many, ...''  but leaves us
wondering how to extract the actual single centered black hole
entropy from (\ref{indexdef}).

It has generally been assumed that  a single centered black hole
dominates the entropy of its charge sector, so the existence of
multicentered   configurations is merely a minor nuisance,
completely negligible in the thermodynamic limit. Indeed when
several black holes are dynamically merged, the second law of
thermodynamics implies that their horizon entropy increases.
However, for our ``molecular''  bound states, merging the centers
  requires adding enough energy to the system to overcome
potential barriers,   necessarily producing a \emph{nonextremal}
final black hole. The possibility remains that multicentered
extremal solutions might  be more entropic than the corresponding
single centered extremal hole.

One of the surprising results reported in \cite{Denef:2007vg} is
that in fact, in suitable parameter regimes, multicentered entropy
\emph{does} dominate  single centered entropy in the uniform large
charge limit. More precisely, when  charges $\Gamma$ obtained by
wrapping D4, D2 and D0 branes around various cycles of $X$ are
scaled up as $\Gamma \to \Lambda \Gamma$, there exist two centered
solutions with   horizon entropy scaling as $\Lambda^3$ while   the
single centered entropy  scales as $\Lambda^2$.

Let us give a  concrete example of this phenomenon, referring to
\cite{Denef:2007vg}  for  more details. Consider type IIA string
theory compactified on $X = T^2_1 \times T^2_2 \times T^2_3$, a
product of three two-tori.  Let $D$ be the 4-cycle $(T^2_1 \times
T^2_2) + (T^2_2 \times T^2_3) + (T^2_3 \times T^2_1)$ and let
$\tilde{D}$ be the 2-cycle $T^2_1+T^2_2+T^2_3$. Then the entropy
function of a charge $\Gamma$ corresponding to $p^0$ D6-branes on
$X$, $p$ D4-branes on $D$, $q$ D2-branes on $\tilde{D}$ and $q_0$
D0-branes is given by
\begin{equation} \label{entropyfunction}
 S(\Gamma) = \pi \sqrt{-4 p^3 q_0 + 3 p^2 q^2 + 6 p^0 p q q_0 - 4 p^0 q^3 - (p^0
 q_0)^2}.
\end{equation}

\EPSFIGURE{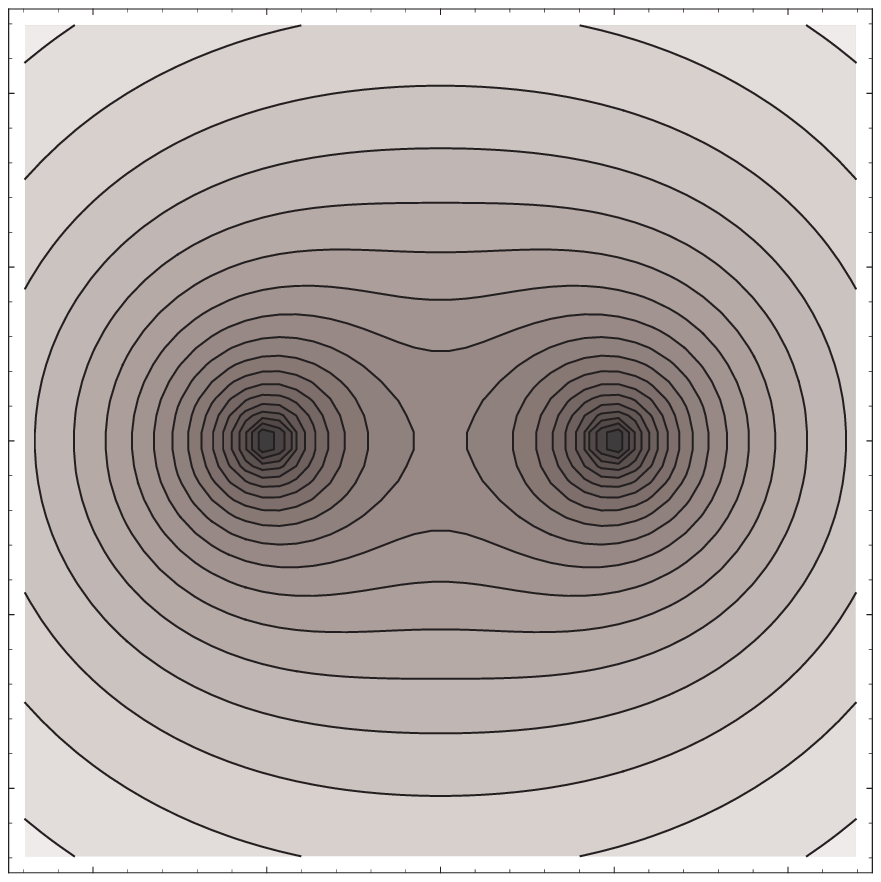,width=5.6cm,angle=0,trim=0 0 0 0}%
{Metric warp factor $\pi/S(H)$ for the 2-centered example.
\label{twocenter}}

\noindent Now consider a total charge
\begin{equation}
\Gamma =
 (p^0,p,q,q_0)=\Lambda(0,6,0,-12),
\end{equation}
in a background in which the area of each
 $T^2$ equals $v$ (which plays the role of $\tau_\infty$). Then
for any  $v$, there exists a single centered solution with horizon
entropy given by (\ref{entropyfunction}):
\begin{equation} \label{singleS}
 S_1 = 72 \sqrt{2} \pi \Lambda^2.
\end{equation}
However, when $v >  \sqrt{18} \Lambda$, there is also a two-centered
bound state with charges
\begin{equation}
 \Gamma_1 = (1,3 \Lambda,6 \Lambda^2,-6 \Lambda),
 \qquad \Gamma_2 = (-1,3 \Lambda,-6 \Lambda^2,-6 \Lambda).
\end{equation}
The constant terms in the harmonic functions (\ref{harmfunct}) are
$h=(0,\frac{1}{\sqrt{2 v}},0,-\sqrt{\frac{v^3}{2}})$, and the
equilibrium separation (\ref{towcentersep}) is $|\vec x_1 - \vec
x_2| = 12 \sqrt{2} \Lambda (9 \Lambda^2-1) \sqrt{v} / (v^2 - 18
\Lambda^2)$. The resulting metric is well defined, with $S(H)$ real
and positive everywhere (see fig.\ \ref{twocenter}). The two centers
have equal horizon entropy, summing up to a total entropy
\begin{equation}
 S_2 = 12 \pi \Lambda \sqrt{3 \Lambda^4 - 1} \sim \Lambda^3,
\end{equation}
which is indeed parametrically larger than the single centered
entropy (\ref{singleS}). (This does not contradict the holographic
principle, since the area of any surface enclosing the black holes
grows at least as fast as $\Lambda^3$.)

When   $v$ is kept fixed while sending $\Lambda \to \infty$,
(\ref{towcentersep}) ceases to have a solution and hence these
2-centered solutions   disappear from the spectrum. However, the
regime in which (\ref{indexdef}) can be reliably computed
microscopically with present technology is precisely the $v \to
\infty$ limit, so we conclude that at large $\Lambda$, the usual
weakly coupled, weakly curved wrapped D-branes are not computing the
entropy of a single black hole, but rather that of complicated
multicentered configurations. The fact that the entropy computations
\cite{Strominger:1996sh,Maldacena:1997de,Vafa:1997gr} matched so
beautifully  to the single black hole entropy is due to the
``accident'' that in the special charge regime in which the
microscopic asymptotics could be extracted, there just happen to be
no competing multicentered solutions. The microscopic counting in
the regime where multi-centered solutions dominate has not yet been
done.

In conclusion, our scholastic question raises a key issue: we need
new ideas to compute microstates in the generic strong coupling
regime. This is \emph{a fortiori} true in the absence of
supersymmetry.

\acknowledgments

This work was supported in part by the DOE under grant
DE-FG02-96ER40949, by the Belgian Federal Office for Scientific,
Technical and Cultural Affairs through the ``Interuniversity
Attraction Poles Programme -- Belgian Science Policy" P5/27 and by
the European Community's Human Potential Programme under contract
MRTN-CT-2004-005104 ``Constituents, fundamental forces and
symmetries of the universe''. FD would like to thank the Galileo
Galilei Institute for Theoretical Physics for hospitality and the
INFN for partial support during the completion of this work.

\end{document}